\documentclass{article}
\usepackage{graphicx} 
\usepackage[a4paper, total={6in, 8in}]{geometry}
\usepackage[sorting=none]{biblatex}
\usepackage[toc,page]{appendix}
\usepackage{ragged2e}
\usepackage{amsmath, amssymb}
\usepackage{color}
\usepackage{booktabs}
\usepackage{array}
\usepackage{floatrow}
\usepackage[label font=bf, 
            labelformat=simple]{subfig}

\newcommand{\absdiv}[1]{%
  \par\addvspace{.5\baselineskip}
  \noindent\textbf{#1}\quad\ignorespaces
}

\bibliography{ref}
\begin{document}
\begin{flushleft}
\textbf{\Large Modelling the impact of improving access to healthcare on Hepatitis B prevalence in the Thai-Myanmar border region}

\bigskip

Anh D. Pham\textsuperscript{a}, 
Robert Moss\textsuperscript{b},
Wirichada Pan-ngum\textsuperscript{c,d},
Rose McGready\textsuperscript{e,f},
Nicholas Geard\textsuperscript{a}
\bigskip

\textbf{a} School of Computing and Information Systems, University of Melbourne, Melbourne, Victoria, Australia
\\
\textbf{b} Melbourne School of Population and Global Health, University of Melbourne, Melbourne, Victoria, Australia
\\
\textbf{c} Mahidol Oxford Tropical Medicine Research Unit, Faculty of Tropical Medicine, Mahidol University, Ratchawithi Rd, Bangkok, 10400, Thailand
\\
\textbf{d} Department of Tropical Hygiene, Faculty of Tropical Medicine, Mahidol University, Ratchawithi Rd, Bangkok, 10400, Thailand
\\
\textbf{e} Shoklo Malaria Research Unit, Mahidol-Oxford Tropical Medicine Research Unit, Faculty of Tropical Medicine, Mahidol University, Mae Sot, Thailand
\\
\textbf{f} Centre for Tropical Medicine and Global Health, Nuffield Department of Medicine, University of Oxford, Oxford, United Kingdom

\end{flushleft}
\vspace{1cm}
\absdiv{Conflict of interest} The authors have no conflict of interest.

\absdiv{Financial support} This study was carried out as part of AP's PhD thesis, which is funded by Melbourne Research Scholarship from the University of Melbourne.

\absdiv{Authors' contributions} All author provided input for the conception of the study. AP implemented the model, conducted experiments, analysed the results and wrote the first draft of the manuscript. NG, RM and WP provided regular supervision and input. All authors read the manuscript, contributed to it by making edits and giving feedback, and approved the submission.
\pagebreak

\begin{abstract}

\absdiv{Introduction}
In Thailand, Hepatitis B is still endemic despite a strong program to eliminate the disease. A higher prevalence is reported in the border region and among migrants due to physical, financial and cultural barriers. Policies and programs targeting the border region and migrant communities have been suggested. Models can be used to understand and quantify the impact of these policies, given they can capture the heterogeneity within the population.

\absdiv{Methods}
In this study, we developed an Agent-based model that captures the differences between the Thai and migrant populations living in this region, notably the higher level of mobility, lower access to healthcare, and the higher prevalence of Hepatitis B among migrants, by modelling the origin of each individual explicitly. We used the model to estimate future trends of Hepatitis B prevalence in Thailand near the border with Myanmar under different scenarios of intervention.

\absdiv{Results}
Our study shows that although the current intervention level is effective in the Thai population, it is insufficient to reach national elimination targets due to high prevalence in migrants. Improving access to healthcare for migrants and the border region could potentially help to reach elimination targets, and we quantified the level of improvement needed to achieve elimination.

\absdiv{Conclusion}
Although there already exist policies to make healthcare more accessible to migrants and the border regions, they are still not yet effective due to financial and cultural barriers. Bringing down those barriers could reduce Hepatitis B prevalence in those communities and regions and contribute to reaching elimination targets in a reasonable timeline.

\absdiv{Keywords}
agent-based modelling – infectious diseases – demographic heterogeneity – Hepatitis B – elimination - simulation
\end{abstract}

\section{Introduction}
\label{intro}
Hepatitis B is the infection of the liver by the hepatitis B virus. Acute Hepatitis B can cause symptoms such as jaundice, vomiting and abdomen pain, while the chronic form often leads to more serious outcomes such as cirrhosis and cancer. Infants and children under 5 years old have a much higher risk of developing a Chronic infection. Hepatitis B is responsible for significant disease burden and death globally \cite{hepbfactsheet}. WHO aims to eliminate Hepatitis B by 2030, with the following elimination targets: 1) reducing new Hepatitis B cases by at least 90\% compared to 2016; 2) reducing Hepatitis B mortality by at leat 65\%; and 3) reducing Hepatitis B prevalence in children under 5 years old to less than 0.1\%.

In Thailand where Hepatitis B is endemic \cite{leroi2016prevalence}, current intervention methods include vaccination, prevention of mother to child transmission (by antiviral treatment and birth dose of the vaccine), and screening the general population and treating those diagnosed with the disease. Thailand started the universal Hepatitis B vaccination program in the period from 1984 to 1992, immunoglobulin and antiviral treatment in 2018 and screening program in 2023. In 2022, Thailand announced the National Strategies to Eliminate Viral Hepatitis, which align its national targets with global goals set by the WHO. 

Despite a strong program to reduce Hepatitis B, a higher prevalence is reported in the Northwestern part of the country due to several reasons. One reason is the incoming flow of migrants and refugees from Myanmar and other countries, where Hepatitis B is often more prevalent than in Thailand \cite{hongjaisee2020prevalence, lee2023refugees} due to low vaccination coverage \cite{anderson2022vaccinationmyanmar}. Moreover, people living in this region, especially migrants, exhibit high mobility \cite{rigg2011Mobility,meemon2021mobility} and frequently relocate over long distances to follow work opportunities or to escape conflicts. This makes it difficult to maintain regular heath check-ups such as antenatal visits, where maternal Hepatitis B can be screened and treated. Finally, access to healthcare is varied between different sites and different cultures, limited by physical barriers and behavioural barriers \cite{khongthanachayopit2017accessibility}. For example, the practice of home birth, while very common in rural Myanmar and Thailand \cite{than2018potential, safebirthreport}, might lead to missing out on the crucial birth-dose of Hepatitis B vaccine especially when a trained attendant is not present. Moreover, although documented migrants are eligible for healthcare services in Thailand for a fee, the majority of migrants are undocumented and thus have little to no access to healthcare \cite{konig2022systematic, 2025myanmarmigrants}.

These factors result in a high prevalence and significant disease burden in this region, and creates pockets of disease that remains largely unaffected by intervention programs and poses a challenge to achieving elimination. \cite{gilder2022hepBvac, posuwan2020towards}. Focusing intervention more on those regions and ethnicities might significantly reduce disease prevalence and burden \cite{banks2016prevalenceborder}. In practice, this might translate to investing in more clinics and personnel in the border region, and adjusting government policy to further support and include migrants in healthcare budget. Conversely, recent funding cuts from host countries of the United Nations threaten to heavily disrupt humanitarian aid worldwide \cite{UNHCR2025fundingcuts}, including healthcare for refugees and migrants living in Thailand-Myanmar border \cite{keay2025uncertainty}. However, the impact of such changes, whether positive or negative, is difficult to measure due to the difficulty in monitoring health and vaccination status of an unstable population. Additionally, due to the main transmission route being vertical from mother to child, a long time horizon is required to observe the results of any changes. Models can help close this gap by providing rough estimates on the long-term outcome of different scenarios. These estimates can be then used to inform policy and optimise intervention strategies.

There have been several modelling studies about Hepatitis B. In a study done in 2016 \cite{nayagam2016requirements}, the authors estimate future status of Hepatitis B worldwide under the current level of interventions as well as scenarios where interventions will be ramped up. The model incorporated the disease progression of Hepatitis B and all routes of transmission as well as demographic and economic data about each country, and suggested that the disease can only be eliminated in more than half of the world within reasonable time frame if interventions are ramped up significantly. Specifically, the requirements include bringing the coverage of infant vaccination, birth-dose vaccination and peripartum antiviral treatment to 90\%, 80\% and 80\% respectively, as well as a population-wide test and treatment rate of 80\%. 
Similar studies targetting specific countries have also made similar suggestions \cite{zou2010hepbchina, mcculloch2020modeling}. In the context of of Thailand, one study investigated the impact of the new screening and treating campaign by the government launched in 2023 \cite{Myka2023hepB}. Most previous studies used mathematical models involving one homogeneous population and did not represent the heterogeneity between different locations and ethnicities within the same country. Meanwhile, as mentioned above, the population in Northwestern Thailand represents a unique challenge, with high immigration and mobility rates \cite{rigg2011Mobility, meemon2021mobility} as well as heterogeneous healthcare access \cite{khongthanachayopit2017accessibility, banks2016prevalenceborder}. To capture the unique dynamics of the population, agent-based models might be more suited due to their innate ability to represent heterogeneity.

According to a recent review, there are very few agent-based models of Hepatitis B, and among which vertical transmission is often not represented \cite{ale2024abmbloodreview}. The most relevant study is perhaps \cite{tian2022highincome}, in which a model was used to investigate the feasibility of eliminating Hepatitis B in Ontario, Canada. Although the model includes several relevant processes such as immigration and vertical transmission, the population was assumed to be relatively stable -- once an immigrant arrives, they are assumed to stay there -- and access to healthcare was assumed to be homogenous. Although these assumptions might be relevant in Canada, they do not apply to Northwestern Thailand. Heterogeneity in healthcare access has been shown to affect the outcome of outbreaks and the endemic level of a disease \cite{patterson2016modeling}. Similarly, despite not often explicitly represented in models of disease transmission in dynamic populations \cite{mogelmose2022incorporating}, internal migration (such as between urban and rural regions in the same country) can strongly influence how a disease spreads. For example, it was shown with mathematical models that if movement from and to an infected subpopulation is possible, then the disease will always remain endemic and reach an equilibrium point \cite{guo2012impactmigrationdisease, gomez2023migrationdisease}.

In this study, we developed a model that captures special characteristics of the population, the disease Hepatitis B and the intervention methods implemented in Thailand. This model was calibrated to match the population as well as the prevalence of Hepatitis B in Thailand at the present, and afterwards used to estimate future trends of Hepatitis B prevalence under different scenarios, specifically answering the following questions: How likely is it for Thailand to achieve a Hepatitis B prevalence of 0.1\% or less in children less than 5 years old with 1) the current level of intervention? 2) changes to access to healthcare for migrants and border region? 3) changes to mobility, treatment and fertility rates? The results show that current intervention level is not sufficient to reach elimination targets if migrants are also taken into account and suggest that health policy in Thailand should do more to reach migrants and rural regions. We also briefly report and discuss the complexity of the model and what it allows us to do.

\section{Methods}
\subsection{The model}
The model needs to represent the different routes of transmission of Hepatitis B, as well as the interventions currently employed and considered in Thailand. It must also capture the unstable and mobile population in border regions, where individuals move around frequently and over long distances. Finally, it needs to represent the difference between regions and ethnicity – for example in remote vs urban, migrants vs Thai – in order to compare between strategies with different focuses.

The model consists of individuals, structured in households and sites, as well as disease progression and transmission. The processes related to population and disease are simulated in discrete timesteps, with each timestep equivalent to one week in real-time.

\subsubsection{Population model}

\begin{figure}[!h]

    \setlength{\labelsep}{0.5mm}
    \centering
    \sidesubfloat[]%
    {
    \includegraphics[width=0.25\textwidth]{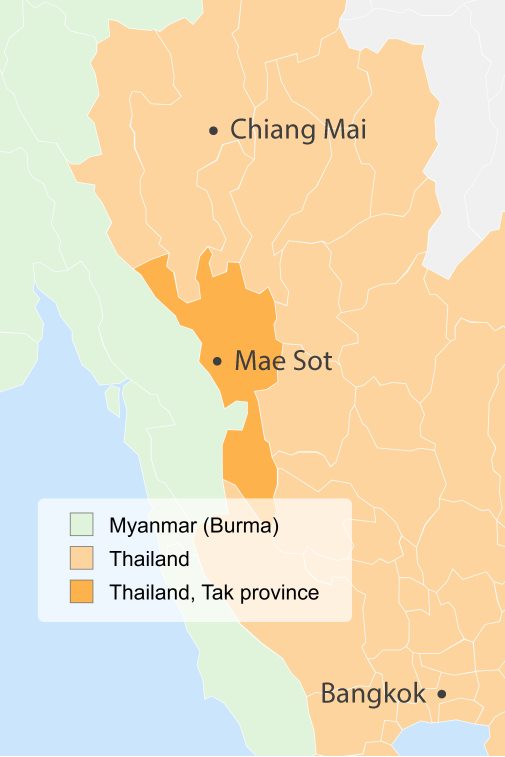}\label{fig:thailand} 
    }
    \hfill
    \sidesubfloat[]%
    {
    \includegraphics[width=0.35\textwidth]{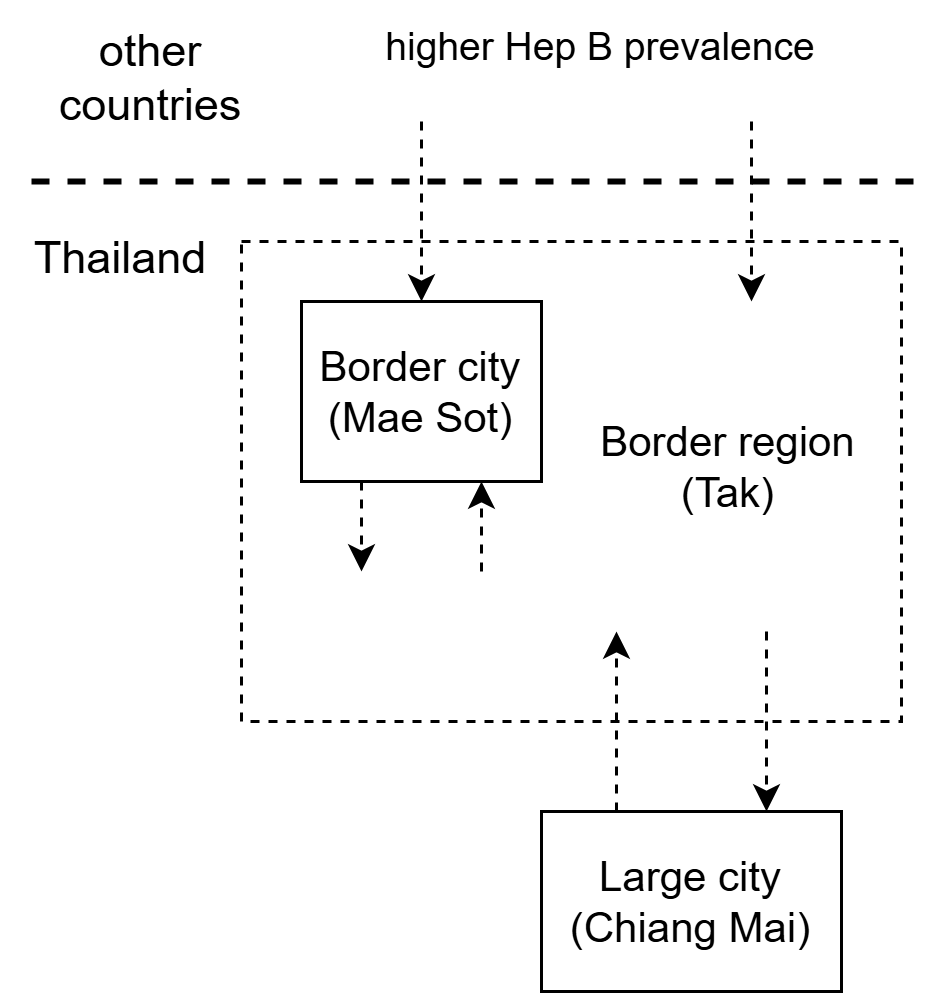}\label{fig:studysites} 
    }  
    \hfill
    \sidesubfloat[]%
    {
    \includegraphics[width=0.25\textwidth]{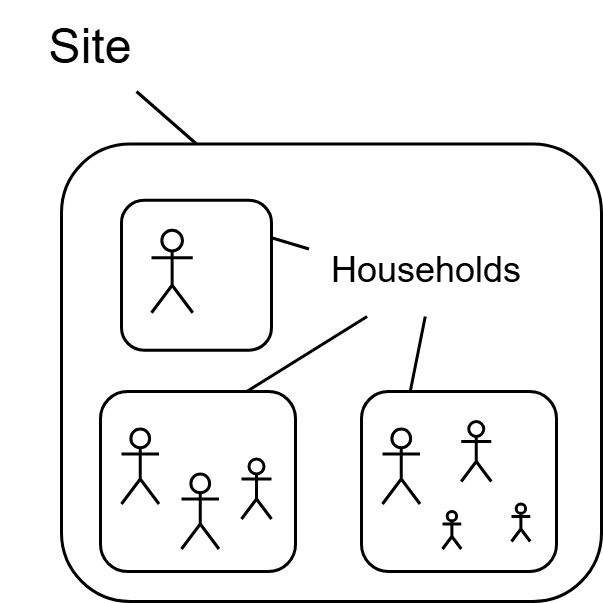}\label{fig:householdsite} 
    }

    \caption{The structure of the population model. \ref{fig:thailand}: the modelled region -- Northwestern border of Thailand. \ref{fig:studysites}: three sites represented in the model, with arrows showing relocation between them. \ref{fig:householdsite}: individuals belong to households, which in turn belong to sites. TODO add some generalisation in discussion about applying it to other contexts}
\end{figure}

The population consists of individuals, grouped into households and sites (Figure \ref{fig:householdsite}). Individuals are characterised by their age, sex and origin (i.e. ethnicity). Individuals age at each timestep, potentially give births and die. Deaths are determined by age- and gender-specific death rates, while births are generated according to the growth rate of the population. In addition to births, new individuals are also added to the model in the form of immigration. Households form and dissolve at each timestep following events such as marriages, divorces, moving out and deaths. Marriages, divorces and moving out events happen with pre-determined probabilities. People in the initial population are assigned a Thai origin, while those who enter through immigration are assigned a Migrant origin. Children are assigned the same origin as their parents. If a child has mixed parents, their origin will be randomly assigned to that of one parent. Death rates and birth rates depend on the origin of an individual. The first generation of migrants -- those who enter the country through immigration -- use Myanmar's growth and death rates instead due to the majority of migrants in this area coming from Myanmar \cite{un2024thaimigrationreport}. Individuals born in Thailand are subjected to Thailand's growth rates and death rates -- including people with migrant origin.

Each household belongs to a site. Sites represent physical areas (such as villages, cities or regions) with a particular level of access to healthcare. We assume that people only interact within their site. At each timestep, households have a probability to move to a different site, modelling the process of relocation. Movement probabilities between different sites are determined by input parameters and also modified by origin -- migrants are assumed to be less stable and thus relocate more often. Other than movement, there are no other interactions between Sites - for example, individuals do not marry individuals from another site. 

The included sites constitute a simplified representation of Northwestern Thailand. They include a small city close to the border (such as Mae Sot), where vaccination and treatment for Hepatitis B are accessible; the surrounding border region (such as Tak province) where healthcare is much less available; and a larger city further away with accessible healthcare and higher density of Thai nationals (such as Chiang Mai or Bangkok). Migrants come from other countries, mainly Myanmar, to the border city and the surrounding region and eventually to the large city (Figure \ref{fig:studysites}).

\subsubsection{Disease model}
\begin{figure}[!h]
    \centering
    \includegraphics[width=0.9\textwidth]{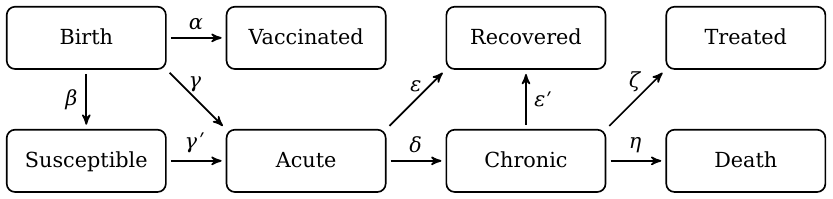}
    \caption{Hepatitis B disease progression from birth.}
    \label{fig:disease}
\end{figure}

Figure \ref{fig:disease} shows the overall progression of the Hepatitis B in our model. The main route of transmission is vertical, from mother to child at birth ($\gamma$). If not infected at birth, individuals enter Susceptible state ($\beta$). While an individual is Susceptible, contacts with infected household members and infected people in the same site can lead to an Acute infection ($\gamma'$). Acute infections can either Recover ($\epsilon$) or lead to a Chronic infection ($\delta$) --  Individuals with chronic infections can sometimes Recover ($\epsilon'$), albeit at a very low rate; undergo Treatment ($\zeta$); or die from the disease ($\eta$).

\paragraph{Horizontal transmission}

The probability of a person being infected ($\gamma'$) horizontally at each time step is given by:
\begin{equation}
    \gamma' = 1 - e^{-foi_{isH}}
\end{equation}

where $foi_{isH}$ is the force of infection on an individual in age class \textit{i}, belonging to a household $H$ \cite{geard2013population}, at site \textit{s}. The force of infection, \textit{foi}, is made up of components of community and household transmission:

\begin{equation} \label{eq:foi}
foi_{isH}=\color{blue}\underbrace{\color{black}q_h\times \frac{I_H}{(N_H-1)}}_{\color{black}\text{Household Transmission}}\color{black}+\color{blue}\underbrace{\color{black}q_c\times \sum \eta_{ij}\frac{I_{j,s}}{N_{j,s}}}_{\color{black}\text{Community Transmission}}
\color{black}
\end{equation}
where:
\begin{itemize}
    \item $q_h=$ Household transmission coefficient
    \item $q_c=$ Community transmission coefficient
    \item $I_H=$ The number of infected individuals in the household
    \item $N_H=$ The number of individuals in the household
    \item $\eta_{ij}=$ Number of daily community contacts between age groups \textit{i} and \textit{j}
    \item $I_{j,s}=$ The number of infected individuals in age group \textit{j} at site \textit{s}
    \item $N_{j,s}=$ The number of individuals in the age group \textit{j} at site \textit{s}
\end{itemize}

\paragraph{Intervention} Interventions include vaccination, prevention of mother-to-child transmission (PMTCT) and screening \& treatment. We assume that children are either Vaccinated ($\alpha$) or not. PMTCT significantly reduces the chance of transmitting the disease from an Acutely or Chronically infected mother to her newborn. Finally, a certain percentage of Chronically infected individuals are screened and Treated at each timestep ($\zeta$).

Access to healthcare varies between sites and between different ethnicities. Both the origin and the site of an individual affect the final probability of getting vaccinated or treated (Eq. \ref{eq:treatprob}).

\begin{equation}
    prob_T(i)=coverage_T \times origin\_access(i) \times site\_access(i)
    \label{eq:treatprob}
\end{equation}

At birth, a child of an uninfected mother can become Vaccinated with the probability determined by Vaccination coverage, or otherwise Susceptible. If the mother is infected with Hepatitis B, the child can also become Infected. In that case, the probability of the child becoming Infected is given by:

\begin{equation}
    \gamma = (1-prob_{PMTCT}(i))\times prob_{MTCT}
\end{equation}
where:
\begin{itemize}
    \item $prob_{PMTCT}(i)=$ probability of receiving effective PMTCT intervention, given in Eq. \ref{eq:treatprob}
    \item $prob_{MTCT}=$ probability of an Infected mother transmitting to her child without intervention (0.95 and 0.7 for Chronically and Acutely infected mothers, respectively)
\end{itemize}

\subsubsection{Data}
Various sources of data are used to parameterise the model. For population model, we used publicly available data such as population growth rate, immigration rate, and age-specific death rates from Thailand and Myanmar \cite{myanmarstats, who2019thailandlifetable, UN2024worldpop}. The specific movement rates between sites are set so that the border city and border region have an even mix of Thai nationals and migrants (approximately 1:1 ratio) while mainly Thai nationals make up the population of the large city (Thai:migrant ratio of approximately 4:1) -- these ratios are derived from personal communication with stakeholders working in the area. In addition, the population is roughly distributed in 1:1:2 ratio for the border city, border region and large city respectively, which roughly resembles the urban:rural population ratio of Thailand (around 55\% of Thai population live in urban areas \cite{desa2018world}) and the size difference between Mae Sot and Chiang Mai. Disease parameters such as vertical transmission rate, probability of developing chronic infection and probability of recovering come from epidemiological studies as well as previous modelling studies in Hepatitis B \cite{juszczyk2000clinical, wen2013mothertoinfant, tian2022highincome}.

Some parameters cannot be found in literature but can be roughly estimates using a relevant statistic. Marriage rate is estimated using percentage of people who are never married in several age-groups \cite{williams2006unmarried}, while divorce rate is estimated using marriage success rate. Specific parameters and data sources can be found in Table \ref{tab:pop_parameters} and Table \ref{tab:dis_params}.

\subsection{Experiments}
\subsubsection{Calibration}
Initially, the population is random generated with some existing data, such as age and household size distribution. The population model is then simulated for 100 years. The purpose is to remove any artifacts from the initial generation so that the synthetic population is more realistic and representative of a real population -- 100 years are adequate for the initial randomly generated population to be completely replaced by a populated created by the dynamic processes in the model. Afterwards, a number of infections are seeded into the population, and the combined population and disease model continued to run for 60 years. As before, the purpose is to allow the disease to spread and realistically distributed among the population and to reach a state of equilibrium. After the total burn-in period of 160 years described above, we start the three intervention methods (vaccination, PMTCT and treatment). This starting point correlates with the year 1984. The model is then run for 40 years to reach the present time. 

For parameters which we cannot find direct values in the literature or relevant statistics to estimate indirectly, we run the model with different values to find one that best fits real data. For the population model, annual moving out probability is not available. We use Thailand's household size distribution from The Household Socio-Economic Survey 2024 to calibrate those parameters. Specifically, we sampled 20 values uniformly distributed in the range [0\%, 1\%] and choose the value that produces the lowest Mean Squared Error against real data. 

In terms of disease transmission,  unknown parameters include horizontal transmission rate and treatment coverage. We use the historical prevalence as well as age-specific prevalence of Hepatitis B in Thailand to calibrate them. Specifically, we sample horizontal transmission coefficient in the range of $[10^{-6}, 10^{-4}]$ and choose the value that, after epidemic burn-in, produces the best fit to prevalence reported in 1984 (around 10\%). Afterwards, we sample annual treatment probability in the range of [1\%, 10\%] and choose the value that produces the best fit to prevalence trajectory from 1984 to 2024 as well as age-specific prevalence reported by some studies 
\cite{leroi2016prevalence, posuwan2016success, nilyanimit2025significant}.

\subsubsection{Likelihood of reaching elimination target with current intervention level}
In this experiment, we investigate the likelihood of reaching a prevalence of 0.1\% or less in children less than 5 years old under the current level of intervention. Specifically, we assume that access to healthcare will remain at the same level, meaning migrants only have 60\% access to vaccination and treatment compared to a Thai person, and people living in border region only have 50\% access compared to those living in the city or village. We run the model for 20 years from now and record prevalence in children under 5 years old over that period. We run 50 simulations in total with different random seeds and aggregate the results.

\subsubsection{Impact of changes to healthcare access}
In the first experiment, we quantify the impact of improving access to healthcare. Firstly, all migrants have 90\% access to healthcare compared to Thai nationals instead of only 60\% (in base scenario). Secondly, healthcare access in the border region is improved to 75\% that of the city and border village instead of 50\% (in base scenario).

In the second experiment, we quantify the impact of reducing access to healthcare in a border city. Specifically, people living in the border city have the same level of access compared to the surrounding border region -- 50\% that of the large city.

In each Scenario, we run the model for 20 years from now and record prevalence in children under 5 years old over that period. We run 50 simulations in total with different random seeds and aggregate the results. In addition to comparing to the elimination target of 0.1\%, we also compare the outcome to the baseline scenario to quantify the impact more concretely.

\begin{table}[!htp]\centering
\caption{Experiment parameters}
\label{tab:pop_parameters}
\scriptsize
\def\arraystretch{2}%
\begin{tabular}{m{0.14\textwidth}m{0.1\textwidth}m{0.11\textwidth}m{0.13\textwidth}m{0.07\textwidth}m{0.09\textwidth}m{0.07\textwidth}m{0.1\textwidth}}\toprule
\textbf{Scenario} &\textbf{Large city access (\%)} &\textbf{Border city access (\%)} &\textbf{Border region access (\%)} &\textbf{Migrant access (\%)} &\textbf{Treatment rate (\%/year)} &\textbf{Mobility level (\%)} &\textbf{Population growth}\\\midrule
Base &100 & 100 & 50 & 60 & 7.5 & 100 & same as 2024 \\\hline
Improved access &100 & 100 & 75 & 90 & 7.5 & 100 & same as 2024\\\hline
Reduced access &100 & 50 & 50 & 60 & 7.5 & 100 & same as 2024 \\\hline
Higher treatment rate &100 & 100 & 50 & 60 & 10.5 & 100 & same as 2024 \\\hline
Reduced mobility &100 & 100 & 50 & 60 & 7.5 & 50 & same as 2024 \\\hline
Falling fertility &100 & 100 & 50 & 60 & 7.5 & 100 & projected by UN World Prospects \\\hline
\bottomrule
\end{tabular}
\end{table}

\subsubsection{Sensitivity analyses: future treatment coverage, mobility and fertility}
In this experiment we explore changes to some model assumptions, such as treatment coverage, mobility in border region and fertility, and their impact on the outcome of the model. There are three scenarios in total: 1) treatment rate will be 40\% higher, 2) mobility will decrease by 50\%, and 3) population (both Thai and migrant populations) will follow projected growth rate instead of the current growth rate. As before, in each scenario, we run the model for 20 years from now and record prevalence in children under 5 years old over that period. We run 50 simulations in total with different random seeds and aggregate the results.

\section{Results}
\subsection{Calibration}

We found that a annual rate of leaving home of 0.025\% produces the best fit in terms of household size distribution to real data. Although there is still a small bias in our model towards smaller households, generally we are able to match the important characteristics of Thailand population quite well in age structure and household size distribution (Figure \ref{fig:popcali}).

\begin{figure}[!htp]
    \centering
        \setlength{\labelsep}{0.5mm}
    \centering
    \sidesubfloat[]%
    {
        \label{fig:diseasecali_prev}
        \includegraphics[width=0.4\textwidth]{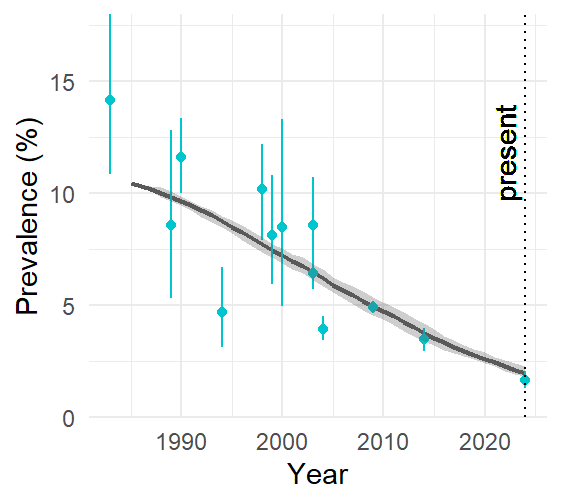}
    }
    \hfill
    \sidesubfloat[]%
    {
        \label{fig:diseasecali_age}
        \includegraphics[width=0.52\textwidth]{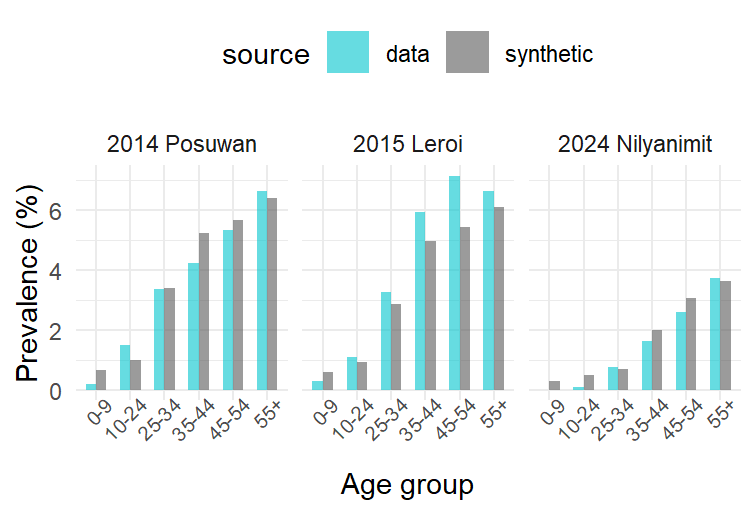}
    }  
    \hfill

    \caption{The synthetic prevalence trends match historical data. In \ref{fig:diseasecali_prev}, red dots represent Hepatitis B prevalence reported in seroprevalence studies, and the error bars represent the 90\% Confidence Interval of each data point. In \ref{fig:diseasecali_age}, red bars represent age-specific prevalence reported in several studies, and grey bars represent the prevalence observed in the model.}
    \label{fig:diseasecali}
\end{figure}

In terms of disease, we found that an annual treatment rate of 7.5\% would produce the best fit to data. Figure \ref{fig:diseasecali} shows that the prevalence trajectory in our model somewhat matches historical prevalence in Thailand. The notable level of discrepancy in historical prevalence reported by previous studies may have been caused by differences in the cohorts and methodology. For example, the 1985 study by Brown et al \cite{brown1989hepatitis} was conducted in leprosy resettlement villages, which were suggested to have a slightly higher Hepatitis B prevalence compared to the general population, while the 2006 study by Chongsrisawat et al \cite{chongsrisawat2006hepatitis} was conducted with a younger cohort compared to Thailand's real age distribution. In addition to prevalence trajectory, the model reproduces similar trends in age-specific prevalence to those reported in Thailand \cite{leroi2016prevalence, posuwan2016success, nilyanimit2025significant}.

\subsection{Likelihood of reaching elimination target with current intervention level}
According to our model, the current level of intervention is unlikely to be sufficient to achieve the elimination target within the next 20 years. Specifically, if only Thai children 0-5 years old are considered, prevalence will reach 0.1\% in 2035; however, prevalence among both migrant and Thai children will only drop to 0.33\% in 2045 (Figure \ref{fig:eli_results}). In terms of sites, even the large city with access to healthcare and majority Thai population would only reach the target from 2040.

\subsection{Impact of changes to healthcare access}
\begin{figure}[!htp]
    \centering
    \includegraphics[scale=0.8]{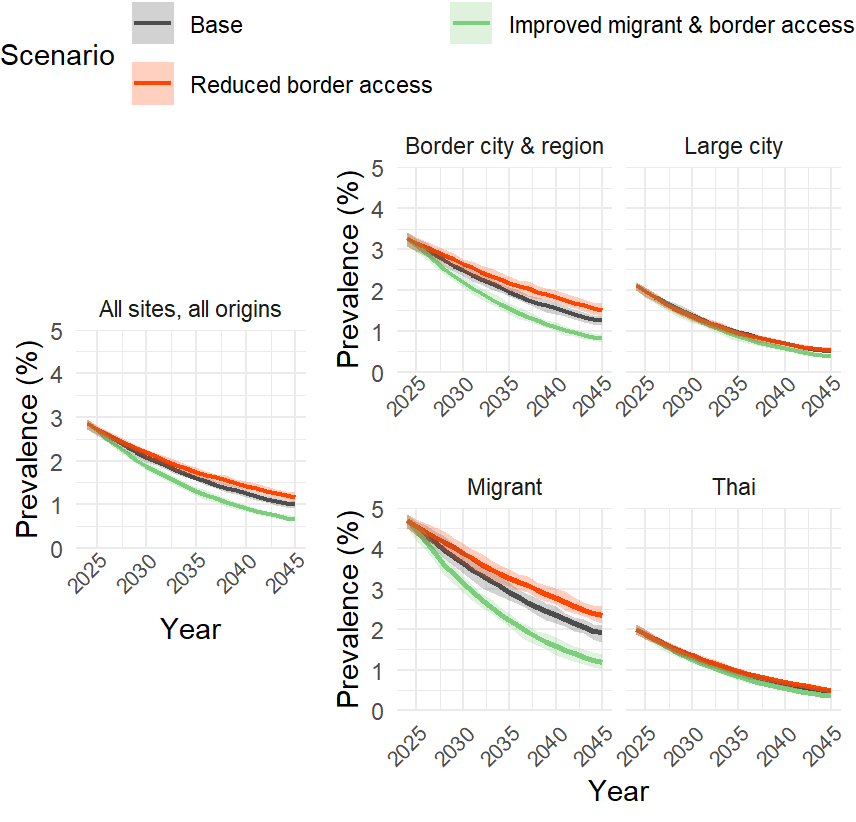}
    \caption{Projected prevalence of Hepatitis B in the populations, 20 years from now, of different sites and different origins. \\ \small{50 simulations were run for each scenario. The central lines represent the mean prevalence in those 50 simulations, while the shaded regions represent the range of prevalence within 90\% Confidence Interval.}}
    \label{fig:site_results}
\end{figure}






\begin{figure}[!htp]
    \centering
    \includegraphics[scale=0.9]{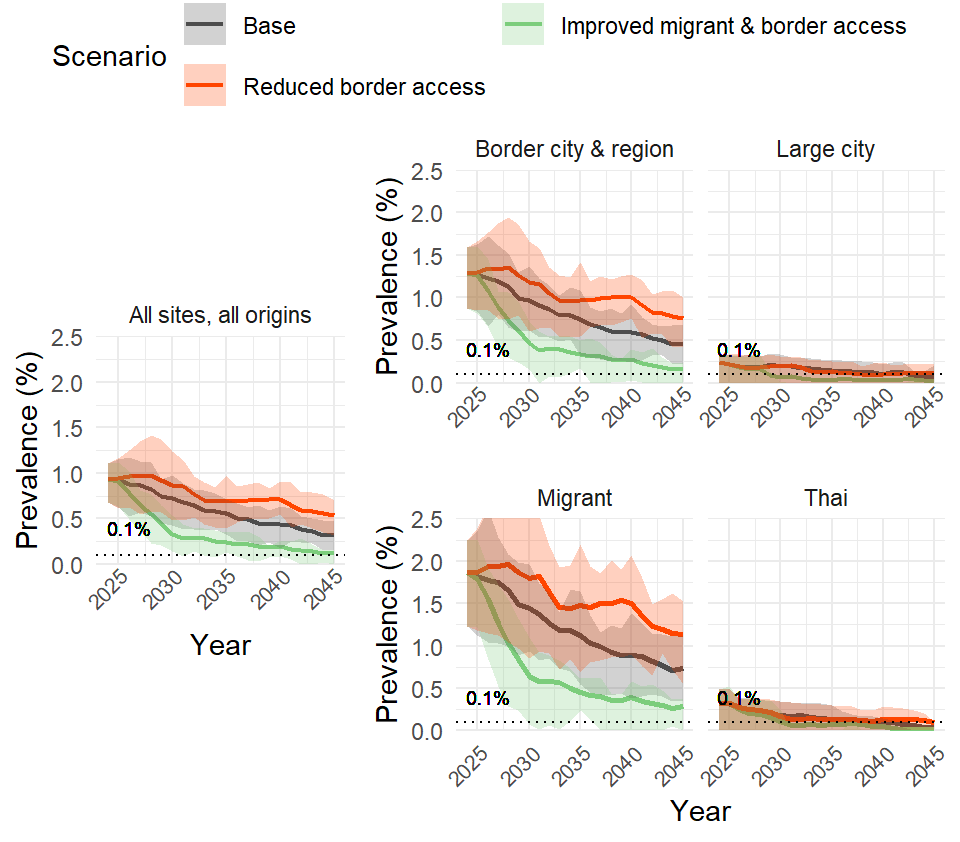}
    \caption{Projected prevalence of Hepatitis B in children 0-5 years old, 20 years from now, of different sites and different origins. \\ \small{50 simulations were run for each scenario. The central lines represent the mean prevalence in those 50 simulations, while the shaded regions represent the range of prevalence within 90\% Confidence Interval.}}
    \label{fig:eli_results}
\end{figure}

Improving access for migrants and border region (green lines in Figure \ref{fig:site_results} and Figure \ref{fig:eli_results}) shows some impact when compared to the base scenario. Particularly, prevalence in the general population after 20 years falls to 0.64\% from 1\% in the base scenario. However, the impact is much more significant when looking at prevalence among children 0-5 years old. Specifically, the elimination target can be achieved in the whole population, with Hepatitis B prevalence among 0-5 year-olds in the large city reaching the 0.1\% mark around 2045. Even in the border city and region (which have a high number of incoming migrants), this figure is reduced to very close to elimination target (0.14\%) compared to the baseline scenario. Among Thai children under 5 years old, prevalence will drop to 0.1\% in 2030 and then further to almost 0\% in 2040.

Unsurprisingly, reducing healthcare access has almost the opposite effect, making reaching elimination target extremely unlikely. By the end of the period, prevalence among children in all sites and of both origins will be around 0.52\%, almost twice that of the base scenario.

\subsection{Impact of potential changes in population, mobility and treatment rate}

\begin{figure}[!htp]
    \centering
    \includegraphics[scale=0.9]{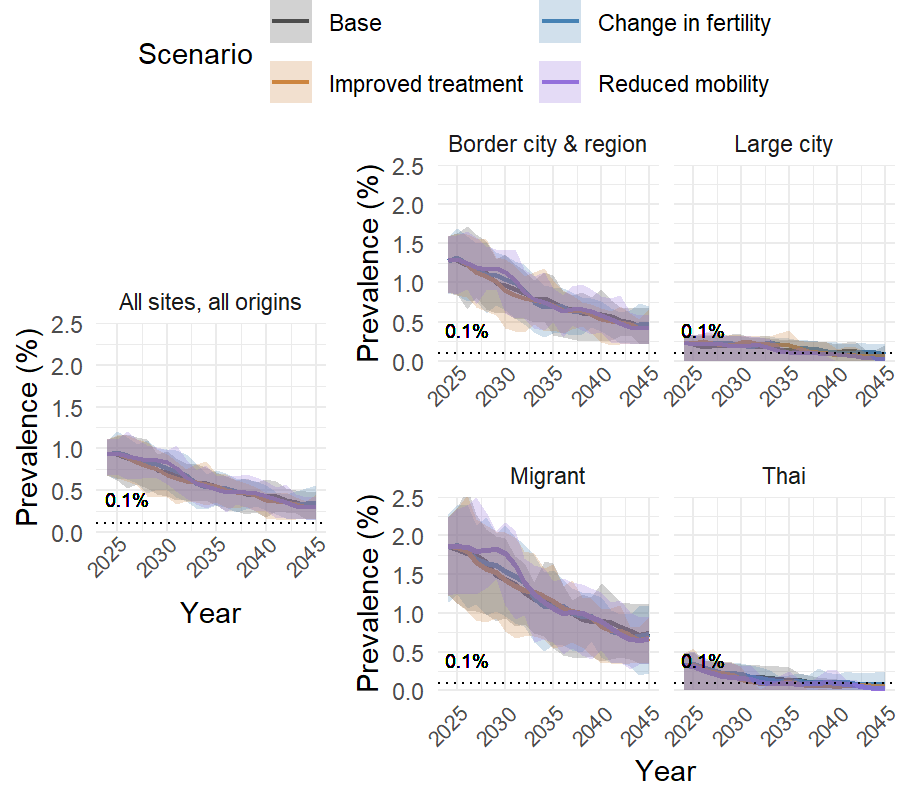}
    \caption{Prevalence of Hepatitis B in children 0-5 years old in simulated population, 20 years from now, under different assumptions in population growth, mobility and treatment. \\ \small{50 simulations were run for each scenario. The central lines represent the mean prevalence in those 50 simulations, while the shaded regions represent the range of prevalence within 90\% Confidence Interval.}}
    \label{fig:sens_results}
\end{figure}

Figure \ref{fig:sens_results} show that outcomes of the model, in terms of prevalence among children 0-5 years old, do not vary greatly when assuming a decreasing natural growth rate in the near future. Similarly, increasing treatment rate by 40\% and reducing mobility by 50\% both do not have a significant impact on children prevalence.

\section{Discussion}
This study explored the impact of public health policies that improve healthcare access for migrants and remote regions in Northwestern Thailand, using an agent-based model adapted to the unique characteristics of this region.

Our results show that HBV prevalence among Thai infant and children less than 5 years old can be reduced to to less than 0.1\% by 2035 under the current level of intervention. However, if migrants are taken into account, under the assumption that the majority of them will settle in Thailand after some time, current level of intervention will not be enough to reach that target even by 2045.

By focusing on the healthcare access of this group as well as the border regions, the situation could be improved significantly. Specifically, if migrants can assess healthcare at a level close to that of Thai nationals and border regions have more access to healthcare, then the target children prevalence can be reached. In that scenario, even if there remain sites with higher prevalence than the target, the gap is still narrowed significantly and elimination is much more likely. We also explored one potential scenario where access to healthcare is reduced in the border city, potentially due to reduced funding. Our results show that it leads to much higher prevalence in children less than 5 years old, in some sites twice as high compared to the baseline scenario. This can potentially undo a lot of the progress made in recent years. Our sensitivity analysis shows that some potential future changes, such as in treatment rate, mobility and population growth, have little impact on prevalence among children 0-5 years old.

Our results are consistent with many modelling studies of Hepatitis B \cite{Myka2023hepB, nayagam2016requirements} and confirm that increasing access to health interventions such as vaccine, PMTCT and screening followed by treatment would be required to achieve some of the target elimination goals. Compare to previous studies, we model origin explicitly and assume that people of migrant origin have less access to healthcare compared to those of Thai origin. This leads to a gap in the Hepatitis B prevalence between Thai nationals and migrants, as well as higher prevalence and a longer timeline to reach elimination targets. Due to continued migration from neighbouring countries and Thailand economy's reliance on migrants \cite{un2024thaimigrationreport, united2020asia}, we believe our model provides estimates that more closely match reality compared to studies that mainly focus on Thai people.

Our findings have several implications on public health. Firstly, prevalence varies greatly between urban and remote sites, and Thai and people of other origins. As such, depending on the selection method, studies can give biased results regarding Hepatitis B status. In terms of public health, although there are already some policies and programs to make healthcare more accessible to migrants and the border regions, they are still not yet effective due to financial and cultural barriers \cite{konig2022systematic}. Bringing down those barriers could reduce Hepatitis B prevalence in those communities and regions and contribute to reaching elimination targets in a reasonable timeline. As mentioned in Section \ref{intro}, in practice this can mean reducing financial costs of healthcare services \cite{hanvoravongchai2025assessing}, adopting and promoting cost-effective treatment \cite{janekrongtham2023cost}, or training home birth attendants and midwives to make sure vaccines are consistently administered (similar to Safe Birth for All \cite{safebirthreport}). Conversely, significant funding cuts \cite{UNHCR2025fundingcuts}, especially in border cities where migrants -- the most vulnerable group -- account for a large portion of the population, are greatly detrimental to the progress towards elimination of Hepatitis B. Finally, we predict that the recently launched screening and treating campaign will not have significant impact on children prevalence in the near future, and so will changes to population growth and mobility.

In this study, modelling origin and site explicitly allows us to quantify the impact of public health policies that target migrants and remote regions, as well as future changes to mobility. Previously, this level of complexity was not often explored in agent-based modelling of Hepatitis B. Moreover, by calibrating the model to closely match previous and current Hepatitis B prevalence in Thailand, we ensure that the model produces estimates that are reasonably reliable. Finally, modelling Hepatitis B with an Agent-based model allows us to incorporate age-dependent chronic risk as well as vertical transmission rates directly from clinical studies.

There are several limitations to our study. Firstly, we made several assumptions that might influence the outcome. We assume that vaccination is fully effective -- if a child is vaccinated, they remains vaccinated for the rest of their life -- and that all people not vaccinated at birth do not attend catch-up vaccination. In reality, protection from vaccines often reduce with time, and people who work in certain fields might get a booster or catch-up vaccination. However, we believe that these assumptions have little impact on prevalence of the chronic state, firstly because they potentially cancel out each other and especially because adults have a very low chance to develop chronic Hepatitis B. We also assume that some demographic parameters, such as death rate or marriage rate, would remain the same in the future. This might not be the case and might change the outlook of the disease. In addition to those assumptions, some modelling decisions are also important. We model movement between sites as a random process with certain rates, while in reality this is more complex and involves individuals making decisions based on factors such as work opportunity, healthcare access and military conflicts. We do not model sexual transmission explicitly, instead we model household transmission and a small degree of community transmission as a proxy. Whether this provides a good approximation is to be investigated.

\section{Acknowledgements}

\printbibliography

\appendix
\makeatletter
\renewcommand \thesection{S\@arabic\c@section}
\renewcommand\thetable{S\@arabic\c@table}
\renewcommand \thefigure{S\@arabic\c@figure}
\makeatother
\setcounter{figure}{0}
\setcounter{table}{0}
\section{Extra figures and tables}
\begin{table}[!htp]\centering
\caption{Population model parameters}
\label{tab:pop_parameters}
\scriptsize
\def\arraystretch{2}%
\begin{tabular}{m{0.25\textwidth}m{0.5\textwidth}m{0.1\textwidth}}\toprule
\textbf{Parameter} &\textbf{Description} &\textbf{Values \& sources} \\\midrule
Population growth rate &Annually, both through natural growth and immigration. & UN World population prospects \cite{UN2024worldpop} \\\hline
Immigration rate &Annual rate of change in population size due to immigration. & UN World population prospects \cite{UN2024worldpop} \\\hline
Age-specific fertility probability& Probability distribution of age of the mother of a birth& Obtained by fertility rates \\\hline
Death rates &Annual Age- and sex-specific probabilities of death by sex and year of age. &Thailand life table, WHO \cite{who2019thailandlifetable} \\\hline
Couple formation parameters &Age range that a currently single individual is eligible to form a couple, and annual probability that this will occur. &(18, 60), 9\% \\\hline
Partner age difference (mean, standard deviation) &Parameters governing the age difference between partners during couple formation. &2, 2 \\\hline
Couple dissolution parameters &Age range that a currently coupled individual is eligible dissolve that couple, and annual probability that this will occur. &(18, 60), 0.5\% \\\hline
Leaving home parameters &Minimum age at which an individual currently living with a parent/guardian will form a new single-person household, and annual probability that this will occur. &18, 0.25\% \\
\bottomrule
\end{tabular}
\end{table}

\begin{table}[!htp]\centering
\caption{Disease model parameters}\label{tab:dis_params}
\scriptsize
\def\arraystretch{2}%
\begin{tabular}{m{0.15\textwidth}m{0.5\textwidth}m{0.2\textwidth}}\toprule
\textbf{Parameter} &\textbf{Description} &\textbf{Values \& sources} \\\midrule
$\gamma$ &Mother-to-child transmission probability & 90\% \cite{gentile2014vertical} \\\hline
$\delta$ & probability of developing a chronic infection from an acute infection & varies by age, from 90\% (newborns) to 10\% (adults) \cite{hyams1995risks} \\\hline
$\epsilon$ & probability of recovering from an acute infection & $1 - \delta$\\\hline
$\epsilon'$ & annual probability of recovering from a chronic infection & 1\% \cite{da1996hepbclearance}\\\hline
$\eta$ & annual probability of dying from a chronic infection & 0.7\% \cite{tian2022highincome}\\\hline
$q_h, q_c$ &Horizontal transmission coefficients & $q_h$: $10^{-4}$ (assumed), $q_c$: $4.7*10^{-6}$ (calibrated) \\\hline
$coverage_{vaccine}$ & ideal vaccination coverage (Thai nationals in a large city) &98\% (assumed) \\\hline
$coverage_{PMTCT}$ &ideal antiviral treatment coverage for pregnant women with Hep B (Thai nationals in a large city) &80\% (assumed) \\\hline
$coverage_{treatment}$ &ideal annual treatment coverage (Thai nationals in a large city) &7.5\% (calibrated)\\\hline
\bottomrule
\end{tabular}
\end{table}

\begin{figure}[!htp]
    \centering
    \includegraphics[scale=0.8]{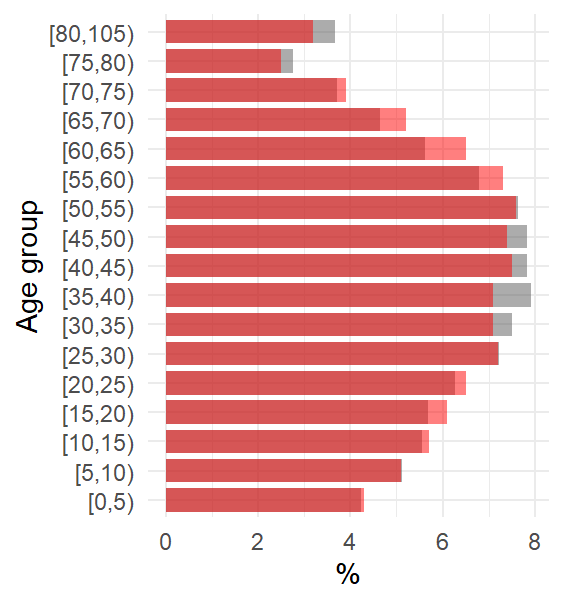}
    \includegraphics[scale=0.8]{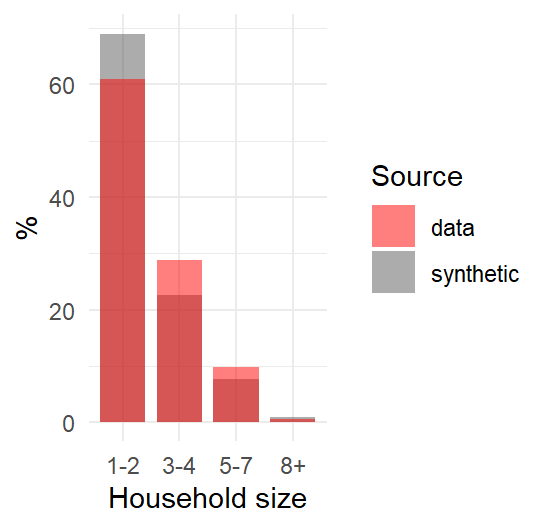}
    \caption{The synthetic population compared to Thailand population}
    \label{fig:popcali}
\end{figure}

\end{document}